\documentclass[aip,apl,reprint,amsmath,amssymb,superscriptaddress,]{revtex4-2}

\usepackage{graphicx}
\usepackage{bm}
\usepackage{dcolumn}
\usepackage{color}
\usepackage{hyperref}
\hypersetup{hidelinks}

\providecommand{\dd}{\mathop{}\!\mathrm{d}}
\newcommand{\im}{\mathrm{Im}}

\newcommand{\rect}{\operatorname{rect}}

\begin{document}

\title{Fixed-detector tilt--defocus sensing by upstream source coding in a time-reversed Young interferometer}

\author{Jianming Wen}
\email{jwen7@binghamton.edu}
\affiliation{Department of Electrical and Computer Engineering, Binghamton University, State University of New York, Binghamton, New York 13902, USA}

\date{\today}

\begin{abstract}
We propose a physically explicit sensing application of a time-reversed Young (TRY) interferometer: simultaneous monitoring of beam tilt and focus drift with a fixed detector. The task is relevant to compact optical relays, free-space links, fiber-coupling stages, and micro-optical alignment modules, where continuous tracking of pointing and focus is needed but downstream wavefront cameras or multiport analyzers are undesirable. Using a finite-width double-slit Fresnel model, we derive the exact local TRY response functions for tilt-like and defocus-like phase perturbations and compute the corresponding optimal upstream source codes numerically. The physical optimal codes are fringe-locked and differ qualitatively from the simple odd/even modes suggested by Gaussian toy models. Two source-coded scalar channels recover essentially all local Fisher information in the full source-resolved TRY record for the physical model considered here. Compared with downstream direct intensity sensing, TRY provides first-order access to the mixed tilt--defocus task with fixed detection; compared with ideal downstream matched-mode sorting, its advantage is architectural rather than fundamental.
\end{abstract}

\maketitle

\emph{Introduction.---} A common practical problem in precision optics is to track \emph{beam steering and focus drift} without resorting to a downstream camera, a Shack--Hartmann array, or a reconfigurable multiport analyzer~\cite{Platt2001,KanaiAPL2008,Rahmani2024,ZhouAPL2023}. This need arises in compact relay optics, free-space optical links, fiber-coupling stages, and alignment-sensitive micro-optical modules~\cite{Kaymak2018,Karioja2000,Ishikawa2003}, where one often wants a low-channel, continuously operating diagnostic that can separate pointing errors from focus errors while keeping the detection hardware simple. Recent work on single-pixel, nanoscale-aperture, and single-shot wavefront-sensing architectures further illustrates the applied interest in compact task-specific optical diagnostics
\cite{LiuAPL2019,PelzmanAPL2019,ZhangAPL2023}.

The time-reversed Young (TRY) interferometer~\cite{Wen2025,WenHybrid,Wen2026a,Wen2026b,Wen2026c,Wen2026d} offers a natural platform for such a task. In TRY, the informative structure is recovered not by scanning a downstream observation plane, but through a source-conditioned response measured with a fixed detector \cite{Wen2025,WenHybrid}. Recent work established deterministic diffraction-free fringes~\cite{Wen2025}, a hybrid source-projector formulation~\cite{WenHybrid}, differential source-basis encoding~\cite{Wen2026a}, and exact null-constrained sensing~\cite{Wen2026b} in this geometry~\cite{Wen2026c,Wen2026d}. These results suggest a broader viewpoint: in TRY, the source coordinate is not merely a scan variable but a programmable measurement basis.

Here we turn that viewpoint into a concrete sensing problem. We ask whether the source basis can directly analyze two practically important aberration modes---tilt and defocus---so that the detector only needs to read a few scalar outputs. The contribution is not a new estimation formalism. The underlying Fisher-information and matched-filter machinery is standard~\cite{Tsang2019,Tsang2016,Rehacek2018}. The new element is the \emph{physical realization}: starting from a finite-width double-slit Fresnel model, we derive the exact local TRY score functions for tilt and defocus and show that the optimal upstream source codes are determined by the real TRY fringe structure. They are strongly oscillatory, quantitatively different from the usual odd/even toy intuition, and sufficient to compress the full local information of the source-resolved TRY record into two fixed-detector channels.

This paper makes three points. First, a physical TRY double-slit system can act as a fixed-detector sensor for simultaneous tilt and focus drift. Second, the true optimal source codes are dictated by the exact Fresnel scores rather than by a Gaussian surrogate. Third, TRY should be understood as an \emph{upstream} modal-analysis architecture: it can outperform plain downstream direct intensity sensing for this mixed local task, while an ideal downstream matched-mode sorter remains an information-theoretic benchmark that can match the same local bound in principle. 

\emph{Physical model.---} We consider a monochromatic scalar field of wavelength $\lambda$, a source coordinate $y$, a double slit in the plane $z=0$, and a fixed detector at transverse coordinate $X_D$~\cite{Wen2025}. The source-to-slit and slit-to-detector distances are $L_1$ and $L_2$. The slit transmission is
\begin{eqnarray}
T(x)=\rect\!\left(\frac{x+d/2}{a}\right)+\rect\!\left(\frac{x-d/2}{a}\right),
\end{eqnarray}
where $d$ is the slit separation and $a$ is the slit width. To model the sensing task explicitly, we place a small aberration at the slit plane,
\begin{eqnarray}
\psi(x;\theta_t,\theta_f)=\theta_t\frac{x}{W}+\theta_f\left(\frac{x}{W}\right)^2,
\quad W=\frac d2,\label{eq:aberration}
\end{eqnarray}
where $\theta_t$ is a tilt-like coefficient and $\theta_f$ is a defocus-like coefficient~\cite{PellatFinet1994,Noll1976}. The normalized coordinate $x/W$ makes both parameters dimensionless and ties them directly to the slit scale.

The detector field is
\begin{eqnarray}
E(y|\theta_t,\theta_f)=\int_{-\infty}^{\infty}\dd x\, T(x)e^{\frac{ik(x-y)^2}{2L_1}}
e^{\frac{ik(X_D-x)^2}{2L_2}}
e^{i\psi(x;\theta_t,\theta_f)},\label{eq:field}
\end{eqnarray}
with wavenumber $k=2\pi/\lambda$, and the TRY response is
\begin{eqnarray}
R(y\mid\theta_t,\theta_f)=|E(y\mid\theta_t,\theta_f)|^2.\label{eq:Rdef}
\end{eqnarray}
Equation~(\ref{eq:field}) is the finite-width Fresnel model used throughout. The source coordinate $y$ labels the upstream basis in which the measurement is performed, while $x$ labels the slit-plane coordinate over which the field contributions are coherently summed.

At the operating point $(\theta_t,\theta_f)=(0,0)$, define
\begin{eqnarray}
E_0(y)=E(y\mid0,0),\quad R_0(y)=|E_0(y)|^2.
\end{eqnarray}
Here, $E_0(y)$ is the unaberrated TRY field and $R_0(y)$ is the corresponding baseline source-resolved response. To determine how tilt and defocus perturb this baseline, we introduce the aberration-weighted Fresnel moments
\begin{align}
M_t(y)&=\int\dd x\,T(x)\frac{x}{W}e^{\,ik\frac{(x-y)^2}{2L_1}}e^{\,ik\frac{(X_D-x)^2}{2L_2}},\\
M_f(y)&=\int\dd x\,T(x)\left(\frac{x}{W}\right)^2
e^{\,ik\frac{(x-y)^2}{2L_1}}e^{\,ik\frac{(X_D-x)^2}{2L_2}}.
\end{align}
These moments are the field responses weighted by the generators of the two aberration modes. Physically, $M_t(y)$ measures how a small odd phase ramp across the slit plane is transferred to the detector for each source coordinate, whereas $M_f(y)$ does the same for a small even quadratic phase. They therefore identify the parts of the propagated field that are relevant to tilt and defocus sensing, respectively. In particular, the quadratic weight makes the defocus channel sensitive to the finite slit width; in the narrow-slit limit this even mode becomes nearly common-mode and its first-order signature is (inevitably) suppressed.

Differentiation of Eq.~(\ref{eq:Rdef}) gives the exact local TRY response functions
\begin{align}
g_t(y)&\equiv\left.\partial_{\theta_t}R(y)\right|_0
=-2\,\im\!\big[E_0^*(y)M_t(y)\big],\label{eq:gt}\\
g_f(y)&\equiv\left.\partial_{\theta_f}R(y)\right|_0
=-2\,\im\!\big[E_0^*(y)M_f(y)\big].\label{eq:gf}
\end{align}
These are the exact first-order sensitivity functions of the TRY response. Each is an interference term between the baseline field and the corresponding aberration-weighted moment. The appearance of the imaginary part is important: only the component of the perturbation that is in optical quadrature with the baseline field changes the detected intensity to first order. Thus, $g_t(y)$ and $g_f(y)$ do not merely quantify sensitivity; they show \emph{where} in the source basis each aberration is converted into measurable detector contrast. These functions will later define the optimal source codes, so they are the central objects linking the physical optics of the double-slit system to the estimation strategy.

\emph{Upstream source-coded receiver.---} In the local regime, the TRY response~(\ref{eq:Rdef}) can be linearized around the operating point as
\begin{eqnarray}
R(y\mid\theta_t,\theta_f)\approx R_0(y)+g_t(y)\theta_t+g_f(y)\theta_f.
\label{eq:localexpand}
\end{eqnarray}
This expression is the starting point of the receiver design. It says that, to first order, the unknown parameters appear only through the two score functions $g_t(y)$ and $g_f(y)$ derived above. The source coordinate $y$ therefore carries a structured sensitivity map: some source locations respond mainly to tilt, others mainly to defocus, and the receiver should weight them accordingly.

Assuming shot-noise-limited detection with a small regularizing background floor $B$, we define the local noise weight
\begin{equation}
N(y)=R_0(y)+B.\label{eq:noiseweight}
\end{equation}
The constant $B$ serves only as a stabilizing floor in regions where $R_0(y)$ becomes very small; it does not represent a new physical parameter of interest. Using Eq.~(\ref{eq:noiseweight}), the Fisher matrix of the \emph{full} source-resolved TRY record is
\begin{equation}
F_{\mu\nu}^{\rm full}=\int\dd y\,\frac{g_\mu(y)g_\nu(y)}{N(y)},
\quad\mu,\nu\in\{t,f\}.\label{eq:Ffull}
\end{equation}
Equation~(\ref{eq:Ffull}) is the local information benchmark for this measurement: it is the maximum information available from the source-resolved TRY response itself.

The purpose of source coding is to compress that full record into a few scalar channels without sacrificing the information relevant to the parameters of interest. For a source code $w_m(y)$, we define the coded observable
\begin{equation}
S_m=\int\dd y\,w_m(y)\,R(y\mid\theta_t,\theta_f),
\label{eq:Sm}
\end{equation}
where $R(y\mid\theta_t,\theta_f)$ is the measured TRY response~(\ref{eq:Rdef}) at the unknown parameter values. Thus $S_m$ is not a new optical field; it is a weighted scalar readout formed from the source-resolved TRY signal. Physically, $w_m(y)$ determines which parts of the source basis are emphasized and which are suppressed.

The natural first choice for the source codes is the matched-filter pair
\begin{equation}
w_t^{\rm raw}(y)=\frac{g_t(y)}{N(y)},\quad
w_f^{\rm raw}(y)=\frac{g_f(y)}{N(y)}.
\label{eq:rawcodes}
\end{equation}
These are the direct local score weights: each code emphasizes the source coordinates that are most informative for one parameter relative to the local noise level. In this raw form, however, the two codes need not be orthogonal to the constant background mode or to each other. For stable two-parameter sensing, we therefore orthogonalize them in the noise metric
\begin{equation}
\langle u,v\rangle_N=\int\dd y\,N(y)\,u(y)v(y),
\label{eq:noisemetric}
\end{equation}
where $u(y)$ and $v(y)$ are placeholders of any two source-code functions like the actual raw source codes $w^{\rm raw}_{t}(y)$ and $w^{\rm raw}_{f}(y)$ defined in Eq.~(\ref{eq:rawcodes}). The resulting optimized codes retain sensitivity to the desired parameter channels while suppressing common background and parameter cross-talk. In practice, each signed code can be implemented as a difference of two positive source patterns; that construction is given in the Supplement.

The coded receiver is described by the mean responses
\begin{eqnarray}
\bar S_m=S_m^{(0)}+\sum_{\mu\in\{t,f\}}\!G_{m\mu}\theta_\mu,\;
G_{m\mu}=\int\dd y\,w_m(y)g_\mu(y),\label{eq:Gmatrix}
\end{eqnarray}
where
\[
S_m^{(0)}=\int\dd y\,w_m(y)\,R_0(y)
\]
is the baseline coded output. The matrix $G_{m\mu}$ is the key local transfer matrix: it tells us how strongly each coded channel responds to tilt and defocus. In other words, the source codes turn the distributed source-space sensitivities $g_t(y)$ and $g_f(y)$ into a small set of scalar sensing channels.

The corresponding channel-noise covariance is
\begin{equation}
\Sigma_{mn}=\int\dd y\,N(y)\,w_m(y)w_n(y).
\label{eq:Sigma}
\end{equation}
For the two-channel receiver used here, the local Fisher matrix becomes
\begin{equation}
F^{\rm TRY}_{\rm 2ch}=G^\mathsf{T}\Sigma^{-1}G.
\label{eq:Ftry}
\end{equation}
Equation~(\ref{eq:Ftry}) shows the role of the source codes clearly. The matrix $G$ sets the signal transfer from the physical aberrations to the coded outputs, while $\Sigma$ sets the noise penalty. If the two optimized codes span the same local score space as $\{g_t,g_f\}$, then $F^{\rm TRY}_{\rm 2ch}$ approaches $F^{\rm full}$. The practical goal of this work is precisely to realize that near-lossless compression with only two fixed-detector channels.

This interpretation also clarifies the significance of the section. Equations~(\ref{eq:localexpand})--(\ref{eq:Ftry}) provide the bridge between the physical TRY optics and the sensing architecture. The physical model supplies the exact score functions $g_t(y)$ and $g_f(y)$; the source-coded receiver converts those distributed sensitivities into two scalar observables; and the Fisher matrix in Eq.~(\ref{eq:Ftry}) quantifies how much of the original local information is retained after that compression.

\begin{figure}[t]
\centering
\includegraphics[width=\linewidth]{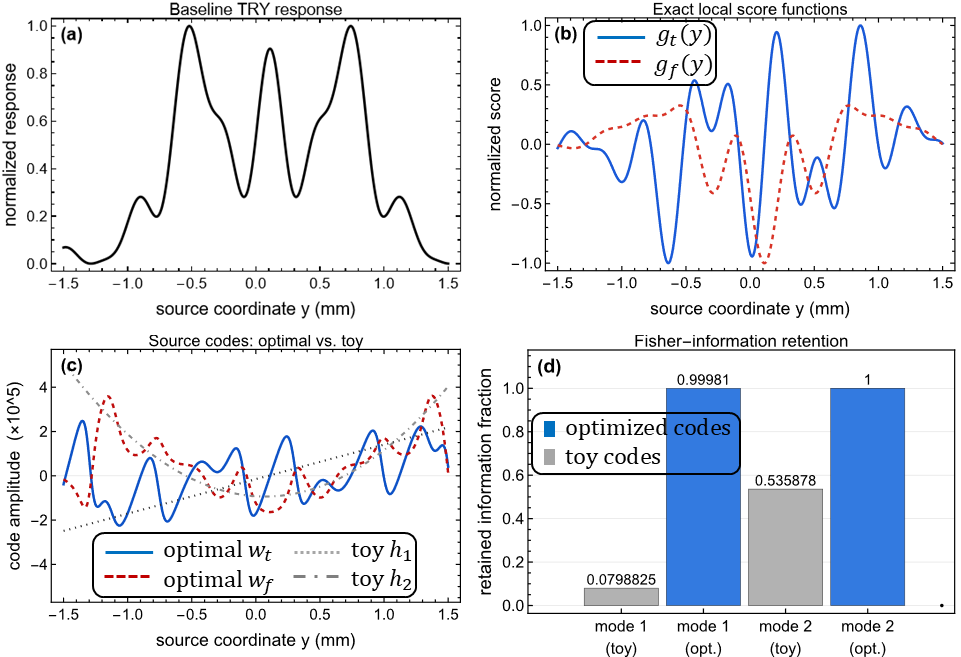}
\caption{Physical TRY response and optimal source codes for the finite-width Fresnel model. (a) Baseline source-resolved TRY response $R_0(y)$. (b) Exact local score functions $g_t(y)$ and $g_f(y)$ for tilt and defocus. (c) Numerically optimized nuisance-orthogonal source codes. The physical optimal codes are oscillatory and locked to the real TRY fringe structure rather than to a smooth odd/even toy basis.}
\label{fig:scores}
\end{figure}

\emph{Numerical Example.---} We now evaluate the physical model and the source-coded receiver described above for a representative visible-wavelength geometry. Unless otherwise stated, we use \(\lambda=633~\mathrm{nm},L_1=L_2=0.35~\mathrm{m},
d=500~\mu\mathrm{m}\), and \(a=250~\mu\mathrm{m}\).
The detector is placed at \(X_D=-L_2\lambda/(4d)\), which positions the working point near quadrature in the corresponding narrow-slit limit and therefore enhances the first-order sensitivity of the TRY response to small phase perturbations. The source coordinate is sampled over \(y\in[-1.5,\,1.5]~\mathrm{mm}\).

Using Eqs.~(\ref{eq:field})--(\ref{eq:Rdef}), we first compute the unaberrated field $E_0(y)$ and baseline response $R_0(y)$. We then evaluate the aberration-weighted Fresnel moments $M_t(y)$ and $M_f(y)$ and obtain the exact local score functions $g_t(y)$ and $g_f(y)$ from Eqs.~(\ref{eq:gt}) and (\ref{eq:gf}). Finally, we form the noise weight
\(N(y)=R_0(y)+B\) with a small regularizing floor
\(B=0.02\,\max_y R_0(y)\), construct the raw matched filters from Eq.~(\ref{eq:rawcodes}), and orthogonalize them in the metric of Eq.~(\ref{eq:noisemetric}). The resulting source codes are then used to build the coded transfer matrix $G$ in Eq.~(\ref{eq:Gmatrix}), the covariance matrix $\Sigma$ in Eq.~(\ref{eq:Sigma}), and the coded Fisher matrix $F_{\rm 2ch}^{\rm TRY}$ in Eq.~(\ref{eq:Ftry}).

Figure~\ref{fig:scores} summarizes the physical behavior of the system. Figure~\ref{fig:scores}(a) shows the baseline source-resolved TRY response $R_0(y)$. Figure~\ref{fig:scores}(b) shows the exact local score functions $g_t(y)$ and $g_f(y)$, which quantify the first-order sensitivities of the TRY response to tilt and defocus at each source coordinate. Figure~\ref{fig:scores}(c) shows the corresponding optimized nuisance-orthogonal source codes. Two features are immediately visible. First, the physical optimal codes are strongly oscillatory and locked to the actual TRY fringe structure. Second, they differ qualitatively from the smooth odd/even forms suggested by the earlier Gaussian surrogate. This difference reflects the fact that the useful information is carried by the true Fresnel-weighted score functions rather than by a generic parity argument alone.

For this physical model, the full Fisher matrix of the source-resolved TRY record [Eq.~(\ref{eq:Ffull})] is
\begin{equation*}
F^{\rm full}=\begin{pmatrix}
5.11999612\times10^{-11} & -6.62496429\times10^{-13}\\
-6.62496429\times10^{-13} & 7.99913250\times10^{-11}
\end{pmatrix},
\end{equation*}
while the optimized two-channel source-coded receiver [Eq.~(\ref{eq:Ftry})] yields
\begin{equation*}
F^{\rm TRY}_{\rm 2ch}=\begin{pmatrix}
5.11939906\times10^{-11} & -6.56482435\times10^{-13}\\
-6.56482435\times10^{-13} & 7.99852673\times10^{-11}
\end{pmatrix}.
\end{equation*}
The two matrices are nearly identical. More quantitatively, the generalized eigenvalues of the whitened retention matrix \((F^{\rm full})^{-1/2}F^{\rm TRY}_{2{\rm ch}}(F^{\rm full})^{-1/2}\) are
\(\{0.99980958,\;1.00000000\}\). Thus two optimized source-coded scalar channels retain essentially all local information available in the full source-resolved TRY measurement for the physical system considered here.

The numerical example also reveals two points that are not evident from the toy-model description. First, the finite slit width is not a minor detail. The even defocus-like mode enters through the quadratic weight in $M_f(y)$, and its first-order visibility depends on sampling that quadratic phase across each slit. In the narrow-slit limit this channel becomes nearly common-mode and the defocus sensitivity collapses. Second, the mismatch between the toy and physical source codes is substantial: when the toy odd/even codes are applied to the same physical system, the retained principal information fractions fall to about $53.7\%$ and $8.0\%$ (Supplement). The physical Fresnel model is therefore essential not only for quantitative accuracy but also for the correct source-code design.

\emph{Comparison with downstream strategies.---} The natural downstream benchmarks for the present problem fall into three classes: single-plane direct intensity sensing, ideal downstream matched-mode sorting, and full-field wavefront analysis~\cite{Platt2001,Tsang2016,Rehacek2018}. The comparison is most meaningful in the same local regime studied above, namely small tilt and defocus perturbations around a fixed operating point.

A \emph{single-plane downstream intensity} measurement is the simplest reference. In that case one records an intensity distribution in an observation plane and estimates the parameters from the resulting image~\cite{Platt2001}. For the mixed tilt--defocus task considered here, however, this is an intrinsically unbalanced receiver: tilt is visible through a first-order shift of the downstream profile, whereas defocus is not generally first-order observable at a single plane without phase diversity or additional wavefront optics~\cite{Paxman1992,Roddier1993,Kendrick1994,Hickson1994,ZhouAPL2023,vanDam2002TipTilt,vanDam2002Defocused,ZhangAPL2023}. In local terms, the corresponding Fisher matrix is therefore rank-deficient at the chosen working point: it captures the tilt channel but leaves the defocus channel strongly suppressed. TRY improves on this baseline by converting both perturbations into first-order measurable contrast in a fixed-detector architecture.

An \emph{ideal downstream matched-mode sorter} is a more demanding benchmark~\cite{Neil2000,Ribak2001}. If the downstream analyzer spans the same local score space as the exact physical TRY scores $\{g_t(y),g_f(y)\}$, then it can in principle attain the same local Fisher matrix~\cite{Tsang2016,Rehacek2018,Hervas2024} as the full source-resolved TRY record. In that sense, TRY does not claim a stronger fundamental information bound. The distinction is instead architectural. In the downstream matched-mode strategy~\cite{Clare2005}, the modal projection is performed after propagation by a dedicated analyzer. In TRY, the same local score extraction is implemented upstream through source coding, while the detector remains fixed.

This difference matters for the low-dimensional tracking problem that motivates this work. TRY replaces a downstream image-plane or modal-analysis problem by an upstream coding problem in the source basis. Once the physical scores are known, the detector hardware is unchanged; recentering, nuisance rejection, and channel updates are implemented by modifying the source codes rather than by redesigning or realigning a downstream analyzer~\cite{Grace2020,LiuAPL2019}. For the physical model considered here, two optimized source-coded channels already recover essentially all local information in the full source-resolved TRY record, as quantified by the near-identity of $F^{\rm TRY}_{\rm 2ch}$ and $F^{\rm full}$ in the numerical example above.

Finally, \emph{full-field downstream wavefront sensing} can provide more information than is needed for the present task~\cite{Platt2001,Kendrick1994,LiuAPL2019,WuAPL2021,ZhouAPL2023,RichterAPL2012,LiuDong2024} because it targets general field reconstruction rather than direct estimation of a few alignment modes. That broader capability comes with higher measurement dimensionality. The role of TRY here is narrower and more specific: it is not proposed as a universal replacement for downstream wavefront sensing, but as a compact fixed-detector alternative when the goal is to monitor a small set of physically meaningful aberration parameters.

Taken together, the comparison is therefore asymmetric but clear. Relative to single-plane direct intensity, TRY provides a genuine local sensing advantage for the mixed tilt--defocus problem. Relative to an ideal downstream matched-mode sorter, the advantage is not fundamental but architectural: TRY relocates the modal analysis upstream into the source basis and achieves the required compression before detection.

\emph{Conclusion---} We developed a physically explicit classical sensing application of a time-reversed Young interferometer: fixed-detector monitoring of tilt and defocus by upstream source coding. Starting from a finite-width double-slit Fresnel model, we derived the exact local TRY response functions for the two aberration modes and computed the corresponding optimal source codes numerically. The physical optimal codes are fringe-aware and differ markedly from the simple odd/even forms suggested by Gaussian toy models. For the system studied here, two optimized source-coded channels retain essentially all local information in the full source-resolved TRY record. Relative to single-plane downstream direct intensity, TRY provides a clear local sensing advantage for the mixed tilt--defocus task; relative to an ideal downstream matched-mode sorter, the advantage is architectural rather than fundamental. These results place TRY in a sharper practical setting as an upstream fixed-detector sensor for low-dimensional alignment and focus-drift monitoring.

\section*{SUPPLEMENTARY MATERIAL}
See supplementary material for the exact score derivation, source-code construction, Fisher-matrix calculation, comparison with Gaussian toy codes, finite-slit-width analysis, and downstream benchmark details.

\begin{acknowledgments}
This work was partially supported by Binghamton University through Startup funds and Watson College internal grant No. 1201479.
\end{acknowledgments}

\section*{Author Declarations}

\subsection*{Conflict of Interest}
The author has no conflicts to disclose.

\subsection*{Author Contributions}
Jianming Wen: Conceptualization; Formal analysis; Methodology; Software; Validation; Visualization; Writing -- original draft; Writing -- review and editing.

\section*{Data Availability}
The data that support the findings of this study are available from the corresponding author upon reasonable request.

\bibliography{references}

\end{document}


\maketitle

\section{Exact local response functions}
This section gives the derivation of the local response functions used in the main text.  The purpose is not to introduce an additional approximation, but to make explicit how the tilt- and defocus-sensitive source-space functions follow from the finite-width Fresnel model.

We write the field at the fixed detector as
\begin{equation}
E(y|\theta_t,\theta_f)=\int_{-\infty}^{\infty}dx\,
T(x)\exp\left[\frac{ik(x-y)^2}{2L_1}+\frac{ik(X_D-x)^2}{2L_2}+i\theta_t\frac{x}{W}+i\theta_f \left(\frac{x}{W}\right)^2\right],\label{eq:S_field}
\end{equation}
where $k=2\pi/\lambda$, $W=d/2$, and $T(x)$ is the finite double-slit transmission.  The source coordinate is $y$, the slit-plane coordinate is $x$, and the detector coordinate $X_D$ is fixed.  Overall Fresnel prefactors that are independent of $x$, $y$, and the unknown parameters are omitted; they only set an overall intensity scale and do not change the source-code structure or the retained information fractions discussed below.

It is useful to separate the unperturbed propagation kernel from the aberration phase.  Define
\begin{equation}
K(x,y)=T(x)\exp\left[\frac{ik(x-y)^2}{2L_1}+\frac{i k(X_D-x)^2}{2L_2}\right],\label{eq:S_kernel}
\end{equation}
and introduce the two dimensionless mode functions
\begin{equation}
q_t(x)=\frac{x}{W},\quad
q_f(x)=\left(\frac{x}{W}\right)^2.\label{eq:S_modes}
\end{equation}
Then Eq.~\eqref{eq:S_field} can be written compactly as
\begin{equation}
E(y|\theta_t,\theta_f)=\int dx\,K(x,y)\,
\exp\left[i\theta_tq_t(x)+i\theta_fq_f(x)\right].
\label{eq:S_compact_field}
\end{equation}
The operating point used throughout the paper is
\begin{equation}
E_0(y)=E(y|0,0)=\int dx\,K(x,y),\quad
R_0(y)=|E_0(y)|^2.\label{eq:S_baseline}
\end{equation}

Because the aperture has finite support and the phase is analytic in $\theta_t$ and $\theta_f$, the parameter derivatives can be taken under the integral sign.  At the operating point,
\begin{equation}
\left.\frac{\partial E}{\partial\theta_\mu}\right|_0
=iM_\mu(y),\quad\mu\in\{t,f\},
\label{eq:S_field_derivative_general}
\end{equation}
where
\begin{equation}
M_\mu(y)=\int dx\,K(x,y)\,q_\mu(x).
\label{eq:S_moment_general}
\end{equation}
Explicitly,
\begin{align}
M_t(y)&=\int dx\,K(x,y)\frac{x}{W},\label{eq:S_Mt}\\
M_f(y)&=\int dx\,K(x,y)\left(\frac{x}{W}\right)^2.
\label{eq:S_Mf}
\end{align}
These are not new fields introduced by a separate optical system.  They are the same Fresnel propagation integral weighted by the slit-plane generators of the two small phase perturbations.

The measured TRY response is the fixed-detector intensity as a function of the upstream source coordinate,
\begin{equation}
R(y|\theta_t,\theta_f)=|E(y|\theta_t,\theta_f)|^2.
\label{eq:S_response}
\end{equation}
For any parameter $\theta_\mu$,
\begin{equation}
\frac{\partial R}{\partial\theta_\mu}=\frac{\partial E}{\partial\theta_\mu}E^*+E\frac{\partial E^*}{\partial\theta_\mu}=2\,\operatorname{Re}\left[
E^*\frac{\partial E}{\partial\theta_\mu}\right].
\label{eq:S_intensity_derivative}
\end{equation}
Using Eq.~\eqref{eq:S_field_derivative_general} at the operating point gives
\begin{equation}
g_\mu(y)\equiv\left.\frac{\partial R(y|\theta_t,\theta_f)}{\partial\theta_\mu}\right|_0
=2\,\operatorname{Re}\left[iE_0^*(y)M_\mu(y)\right]
=-2\,\operatorname{Im}\left[E_0^*(y)M_\mu(y)\right].
\label{eq:S_score_general}
\end{equation}
Thus the two local TRY response functions are
\begin{align}
g_t(y)&=-2\,\operatorname{Im}\left[E_0^*(y)M_t(y)\right],\label{eq:S_gt}\\
g_f(y)&=-2\,\operatorname{Im}\left[E_0^*(y)M_f(y)\right].\label{eq:S_gf}
\end{align}
The minus sign in Eqs.~\eqref{eq:S_gt} and \eqref{eq:S_gf} follows from
$\operatorname{Re}(iz)=-\operatorname{Im}(z)$.  Physically, the first-order intensity change is produced only by the part of the perturbation-weighted field that is in optical quadrature with the unperturbed field.  A perturbation component parallel to $E_0(y)$ changes only the local optical phase and is therefore invisible to intensity at first order.

For clarity, we also state how the term ``score'' is used in the main text.  The functions $g_t(y)$ and $g_f(y)$ are local intensity-response functions,
\begin{equation}
R(y|\theta_t,\theta_f)=R_0(y)+g_t(y)\theta_t+g_f(y)\theta_f+O(\theta^2).\label{eq:S_linear_response}
\end{equation}
When a noise weight $N(y)$ is introduced, the corresponding noise-weighted statistical score directions are proportional to
\begin{equation}
\frac{g_t(y)}{N(y)},\quad\frac{g_f(y)}{N(y)}.
\label{eq:S_weighted_scores}
\end{equation}
These are the raw matched-filter source codes used before nuisance-mode orthogonalization.

A useful consistency check is the narrow-slit limit.  If each slit is approximated by a point at $x=\pm W$, then
\begin{equation}
q_f(+W)=q_f(-W)=1.\label{eq:S_narrow_defocus}
\end{equation}
The defocus-weighted moment then becomes approximately
\begin{equation}
M_f(y)\simeq E_0(y),\label{eq:S_Mf_narrow}
\end{equation}
so that
\begin{equation}
g_f(y)\simeq-2\,\operatorname{Im}\left[ |E_0(y)|^2\right]=0.\label{eq:S_gf_narrow}
\end{equation}
Thus a quadratic phase is almost common-mode in the point-slit limit and has little first-order intensity signature.  Finite slit width is therefore essential for making the even defocus-like mode observable.  By contrast, the tilt generator satisfies $q_t(+W)=1$ and $q_t(-W)=-1$, so it remains a differential phase between the two slit contributions and can produce first-order TRY contrast.

\section{Numerical evaluation and source-code construction}
This section gives the numerical procedure used to construct the source codes shown in the main text.  The calculation uses the same finite-width Fresnel model as Sec.~S1; no Gaussian envelope, point-slit approximation, or parity-based ansatz is imposed.

For the representative geometry, we use
\begin{equation}
\lambda=633~{\rm nm},\quad L_1=L_2=0.35~{\rm m},\quad
d=500~\mu{\rm m},\quad a=250~\mu{\rm m},
\label{eq:S_parameters}
\end{equation}
with
\begin{equation}
W=\frac{d}{2},\quad X_D=-\frac{L_2\lambda}{4d}.
\label{eq:S_detector_position}
\end{equation}
The source coordinate is evaluated over
\begin{equation}
y\in[-1.5,1.5]~{\rm mm}.\label{eq:S_y_window}
\end{equation}
For each value of $y$, the slit-plane integrals are evaluated only over the two finite openings,
\begin{equation}
\int dx\,T(x)F(x)=\sum_{s=\pm 1}\int_{sd/2-a/2}^{sd/2+a/2}dx\,F(x).\label{eq:S_finite_slit_integral}
\end{equation}
Thus the finite slit width enters directly through the numerical integration range rather than through an effective Gaussian or narrow-aperture replacement.

The computed quantities are the baseline field $E_0(y)$, the baseline response $R_0(y)$, the two weighted Fresnel moments $M_t(y)$ and $M_f(y)$, and the two local response functions
\begin{equation}
g_t(y)=-2\,\operatorname{Im}\!\left[E_0^*(y)M_t(y)\right],
\qquad
g_f(y)=-2\,\operatorname{Im}\!\left[E_0^*(y)M_f(y)\right].
\label{eq:S_numerical_scores}
\end{equation}
These functions are real-valued because they are derivatives of the real intensity response.

The noise weight used for the local calculation is
\begin{equation}
N(y)=R_0(y)+B,\quad B=\beta\max_yR_0(y),\quad
\beta=0.02.\label{eq:S_noise_floor}
\end{equation}
The small floor $B$ prevents the inverse-noise weight from becoming numerically unstable near deep minima of $R_0(y)$.  It is not treated as an additional parameter to be estimated.  Because $B$ is chosen proportional to the peak response, an overall rescaling of the optical power rescales both $R_0$ and $N$ consistently and does not change the information-retention ratios reported below.

The natural inner product for source-code construction is the noise-weighted product
\begin{equation}
\langle u,v\rangle_N=\int dy\,N(y)u(y)v(y),\quad
\|u\|_N=\sqrt{\langle u,u\rangle_N}.
\label{eq:S_noise_inner_product}
\end{equation}
With this metric, the single-parameter matched-filter code is proportional to $g_\mu(y)/N(y)$.  This follows directly from the Cauchy--Schwarz inequality: for a code $w(y)$, the squared local signal-to-noise ratio for parameter $\theta_\mu$ is proportional to
\begin{equation}
\frac{\left[\int dy\,w(y)g_\mu(y)\right]^2}{\int dy\,N(y)w^2(y)},\label{eq:S_single_parameter_snr}
\end{equation}
which is maximized by
\begin{equation}
w_\mu(y)\propto\frac{g_\mu(y)}{N(y)}.
\label{eq:S_single_parameter_matched_filter}
\end{equation}
Therefore the raw two-parameter source codes are
\begin{equation}
w_t^{\rm raw}(y)=\frac{g_t(y)}{N(y)},\quad
w_f^{\rm raw}(y)=\frac{g_f(y)}{N(y)}.
\label{eq:S_raw_codes}
\end{equation}

The raw codes are locally optimal score weights, but they need not be orthogonal to the constant source mode or to each other.  We therefore apply a noise-metric Gram--Schmidt procedure. Let $u_0(y)\equiv 1$ denote the constant source-code mode. First, the tilt code is projected away from the constant mode,
\begin{equation}
\widetilde w_t(y)=w_t^{\rm raw}(y)-\frac{\langle w_t^{\rm raw},1\rangle_N}{\langle 1,1\rangle_N}.
\label{eq:S_tilt_projection}
\end{equation}
Next, the defocus code is projected away from both the constant mode and the tilt code,
\begin{equation}
\widetilde w_f(y)=w_f^{\rm raw}(y)-\frac{\langle w_f^{\rm raw},1\rangle_N}{\langle 1,1\rangle_N}-
\frac{\langle w_f^{\rm raw},\widetilde w_t\rangle_N}
{\langle\widetilde w_t,\widetilde w_t\rangle_N}
\widetilde w_t(y).\label{eq:S_defocus_projection}
\end{equation}
Because $\widetilde w_t$ is already orthogonal to the constant mode, Eq.~\eqref{eq:S_defocus_projection} also leaves $\widetilde w_f$ orthogonal to the constant mode.  The order of the Gram--Schmidt construction fixes a convenient channel basis; it does not change the two-dimensional score subspace retained by the receiver.

The normalized source codes are
\begin{equation}
w_t(y)=\frac{\widetilde w_t(y)}{\|\widetilde w_t\|_N},\quad w_f(y)=\frac{\widetilde w_f(y)}{\|\widetilde w_f\|_N}.\label{eq:S_normalized_codes}
\end{equation}
They satisfy
\begin{equation}
\langle w_t,1\rangle_N=0,\quad\langle w_f,1\rangle_N=0,\quad\langle w_t,w_f\rangle_N=0,
\quad\langle w_t,w_t\rangle_N=\langle w_f,w_f\rangle_N=1.\label{eq:S_code_orthonormality}
\end{equation}
The overall sign of each code is arbitrary.  For plotting and interpretation, the signs may be chosen so that the dominant diagonal responses to $\theta_t$ and $\theta_f$ are positive.

In a discretized calculation with source samples $y_j$ and spacing $\Delta y$, the replacement
\begin{equation}
\int dy\,A(y)\;\longrightarrow\;\Delta y\sum_j A(y_j)
\label{eq:S_discrete_integral}
\end{equation}
is used consistently.  For example,
\begin{equation}
\langle u,v\rangle_N\simeq\Delta y\sum_jN(y_j)u(y_j) v(y_j).\label{eq:S_discrete_inner_product}
\end{equation}
The same discretization is used later for the coded response matrix and noise covariance.  This consistency is more important than the particular normalization convention for the codes, because any nonsingular rescaling or rotation of the two coded outputs is accounted for by the transfer matrix and covariance matrix in the Fisher calculation.

Finally, a signed source code does not require a negative optical intensity.  Each code can be decomposed into two nonnegative patterns,
\begin{equation}
w_\mu(y)=w_\mu^+(y)-w_\mu^-(y),\quad
w_\mu^\pm(y)=\max\!\left[\pm w_\mu(y),0\right],
\quad\mu\in\{t,f\}.
\label{eq:S_positive_negative_split}
\end{equation}
The signed coded observable is then obtained from two positive measurements,
\begin{equation}
S_\mu=S_\mu^+-S_\mu^-,\quad S_\mu^\pm=\int dy\,w_\mu^\pm(y)R(y|\theta_t,\theta_f).
\label{eq:S_positive_measurements}
\end{equation}
If the two positive patterns are measured with calibrated exposure factors, the subtraction reproduces the desired signed code.  In the shot-noise model used here, the positive and negative supports are disjoint, so the noise variance of the difference is
\begin{equation}
{\rm Var}(S_\mu^+-S_\mu^-)=\int dy\,N(y) \left[(w_\mu^+)^2+(w_\mu^-)^2\right]=\int dy\,N(y)w_\mu^2(y).\label{eq:S_signed_code_noise}
\end{equation}
Thus the positive-only implementation realizes the same signed-code covariance assumed in the local Fisher analysis, up to known calibration factors.

\section{Full and coded Fisher matrices}
This section explains how the information-retention numbers are computed.  We work in the same local model used in the main text and in Secs.~S1--S2,
\begin{equation}
R(y|\theta_t,\theta_f)=R_0(y)+\sum_{\mu\in\{t,f\}} g_\mu(y)\theta_\mu+O(\theta^2),
\label{eq:S_local_response_for_Fisher}
\end{equation}
with local noise weight
\begin{equation}
N(y)=R_0(y)+B.\label{eq:S_noise_for_Fisher}
\end{equation}
The noise is treated as locally parameter independent.  This is the usual local Fisher approximation for weak perturbations around the chosen operating point.

For the full source-resolved TRY record, the Fisher matrix is
\begin{equation}
F^{\rm full}_{\mu\nu}=\int dy\,
\frac{g_\mu(y)g_\nu(y)}{N(y)},\quad\mu,\nu\in\{t,f\}.
\label{eq:S_full_Fisher}
\end{equation}
This is the benchmark information available before source-code compression.  Equivalently, if we define the noise-weighted score functions
\begin{equation}
s_\mu(y)=\frac{g_\mu(y)}{N(y)},
\label{eq:S_statistical_score}
\end{equation}
then Eq.~\eqref{eq:S_full_Fisher} can be written as
\begin{equation}
F^{\rm full}_{\mu\nu}=\langle s_\mu,s_\nu\rangle_N,
\quad\langle u,v\rangle_N=\int dy\,N(y)u(y)v(y).
\label{eq:S_full_Fisher_inner_product}
\end{equation}
Thus the full Fisher matrix is simply the Gram matrix of the two score directions in the noise-weighted source-code space.

Now consider a set of coded scalar outputs
\begin{equation}
S_m=\int dy\,w_m(y)R(y|\theta_t,\theta_f),
\label{eq:S_coded_output}
\end{equation}
where $m$ labels the source-code channel.  To first order,
\begin{equation}
\overline S_m=S_m^{(0)}+\sum_{\mu\in\{t,f\}}G_{m\mu}\theta_\mu,\label{eq:S_coded_mean}
\end{equation}
with
\begin{equation}
S_m^{(0)}=\int dy\,w_m(y)R_0(y),\quad G_{m\mu}=\int dy\,w_m(y)g_\mu(y).\label{eq:S_coded_transfer}
\end{equation}
The channel-noise covariance is
\begin{equation}
\Sigma_{mn}=\int dy\,N(y)w_m(y)w_n(y)=\langle w_m,w_n\rangle_N.\label{eq:S_coded_covariance}
\end{equation}
The Fisher matrix of the coded receiver is therefore
\begin{equation}
F^{\rm code}=G^{\rm T}\Sigma^{-1}G.
\label{eq:S_coded_Fisher}
\end{equation}
For the two optimized TRY source codes, we denote this matrix by
\begin{equation}
F^{\rm TRY}_{2{\rm ch}}=G^{\rm T}\Sigma^{-1}G.
\label{eq:S_TRY_two_channel_Fisher}
\end{equation}

Equation~\eqref{eq:S_coded_Fisher} has a useful geometric interpretation.  Let
\begin{equation}
\mathcal W=\operatorname{span}\{w_1,w_2,\ldots,w_M\}
\label{eq:S_code_subspace}
\end{equation}
be the source-code subspace, and let $P_{\mathcal W}$ be the orthogonal projector onto this subspace using the noise metric $\langle\cdot,\cdot\rangle_N$.  Then
\begin{equation}
F^{\rm code}_{\mu\nu}=\left\langle P_{\mathcal W}s_\mu,P_{\mathcal W}s_\nu\right\rangle_N.
\label{eq:S_projected_Fisher}
\end{equation}
Thus source coding keeps the components of the full score functions that lie inside the chosen code subspace and discards the orthogonal components.  If the source codes span the same two-dimensional space as the score functions $s_t(y)$ and $s_f(y)$, then
\begin{equation}
F^{\rm code}=F^{\rm full}.
\label{eq:S_no_loss_condition}
\end{equation}
If the span is only approximate, the information loss is exactly the loss of the score components outside $\mathcal W$.

This projection viewpoint also explains why the result does not depend on a particular normalization of the source codes.  Any nonsingular linear recombination of the coded outputs changes $G$ and $\Sigma$, but leaves
\begin{equation}
G^{\rm T}\Sigma^{-1}G \label{eq:S_basis_invariance}
\end{equation}
unchanged.  Therefore the Fisher matrix depends only on the subspace spanned by the codes, not on the particular basis used to represent that subspace.  When the normalized codes of Sec.~S2 are used, $\Sigma$ is the identity matrix to numerical precision; nevertheless, Eq.~\eqref{eq:S_coded_Fisher} is the more general and basis-invariant expression.

For the physical geometry used in the main text, the full source-resolved Fisher matrix is
\begin{equation}
F^{\rm full}=\begin{pmatrix}
5.11999612\times 10^{-11} & -6.62496429\times 10^{-13}\\
-6.62496429\times 10^{-13} & 7.99913250\times 10^{-11}
\end{pmatrix}.\label{eq:S_full_Fisher_numeric}
\end{equation}
Using the two optimized nuisance-orthogonal source codes gives
\begin{equation}
F^{\rm TRY}_{2{\rm ch}}=\begin{pmatrix}
5.11939906\times 10^{-11} & -6.56482435\times 10^{-13}\\
-6.56482435\times 10^{-13} & 7.99852673\times 10^{-11}
\end{pmatrix}.\label{eq:S_TRY_Fisher_numeric}
\end{equation}
The two matrices are nearly identical, showing that the two coded channels span almost the full local score space.

To quantify the retained information in a basis-independent way, we define the information-retention matrix
\begin{equation}
\mathcal R=\left(F^{\rm full}\right)^{-1/2}
F^{\rm TRY}_{2{\rm ch}}\left(F^{\rm full}\right)^{-1/2}.\label{eq:S_retention_matrix}
\end{equation}
The eigenvalues of $\mathcal R$ give the retained Fisher information along the principal parameter directions of the full measurement.  For the numerical example,
\begin{equation}
\operatorname{eig}(\mathcal R)=
\{0.99980958,\;1.00000000\}.
\label{eq:S_retention_eigenvalues}
\end{equation}
The smaller value is the worst retained fraction over all local linear combinations of tilt and defocus.  Therefore, even in the least favorable parameter direction, the optimized two-channel source-coded TRY receiver retains more than $99.98\%$ of the information available in the full source-resolved TRY record.

In the discrete numerical implementation, the same quadrature rule is used for all quantities:
\begin{equation}
F^{\rm full}_{\mu\nu}\simeq\Delta y\sum_j
\frac{g_\mu(y_j)g_\nu(y_j)}{N(y_j)},
\label{eq:S_discrete_full_Fisher}
\end{equation}
\begin{equation}
G_{m\mu}\simeq\Delta y\sum_j w_m(y_j)g_\mu(y_j),
\quad\Sigma_{mn}\simeq\Delta y\sum_j N(y_j)w_m(y_j)w_n(y_j).\label{eq:S_discrete_coded_Fisher}
\end{equation}
Using a consistent discretization for $F^{\rm full}$, $G$, and $\Sigma$ is essential, because the comparison is a statement about information retained by compression rather than about absolute normalization.

\section{Comparison with a Gaussian toy-code basis}
This section explains the quantitative comparison between the physically optimized source codes and the simpler odd/even codes suggested by a Gaussian toy model.  The purpose is not to introduce a competing physical model, but to show why the exact finite-width Fresnel model is needed for source-code design.

In a smooth Gaussian surrogate, a small transverse shift is naturally associated with an odd first-order Hermite-like mode, while a small width or curvature change is associated with an even second-order Hermite-like mode.  This motivates the toy-code intuition
\begin{equation}
h_t(\xi)\sim\xi,\quad h_f(\xi)\sim \frac{\xi^2-1}{\sqrt{2}},\label{eq:S_toy_Hermite_intuition}
\end{equation}
where $\xi$ is a dimensionless source coordinate.  Such codes are useful as a qualitative guide, but they do not include the oscillatory Fresnel phase structure of the physical TRY response.

To make the comparison as fair as possible, the toy codes are not chosen with an arbitrary scale.  Instead, the dimensionless coordinate $\xi$ is defined from the actual baseline TRY response,
\begin{equation}
\xi=\frac{y-\bar y}{\sigma_y},
\label{eq:S_toy_coordinate}
\end{equation}
where
\begin{equation}
\bar y=\frac{\int dy\,y R_0(y)}{\int dy\,R_0(y)},
\quad\sigma_y^2=\frac{\int dy\,(y-\bar y)^2R_0(y)}
{\int dy\,R_0(y)}.\label{eq:S_toy_mean_variance}
\end{equation}
Thus the toy basis uses the centroid and width of the actual physical baseline rather than those of an externally imposed Gaussian.

The raw toy codes are then taken as
\begin{equation}
h_t^{(0)}(y)=\xi,\quad h_f^{(0)}(y)=\frac{\xi^2-1}{\sqrt{2}}.\label{eq:S_raw_toy_codes}
\end{equation}
These functions are only templates.  Before computing Fisher information, they are processed in the same way as the physical matched-filter codes: the constant source mode is removed, the two codes are orthogonalized in the noise metric, and the result is normalized.  Explicitly, with
\begin{equation}
\langle u,v\rangle_N=\int dy\,N(y)u(y)v(y),
\label{eq:S_toy_inner_product}
\end{equation}
and with $u_0(y)\equiv 1$, we define
\begin{equation}
\widetilde h_t(y)=h_t^{(0)}(y)-\frac{\langle h_t^{(0)},u_0\rangle_N}{\langle u_0,u_0\rangle_N}
u_0(y),\label{eq:S_toy_tilt_projection}
\end{equation}
and
\begin{equation}
\widetilde h_f(y)=h_f^{(0)}(y)-\frac{\langle h_f^{(0)},u_0\rangle_N}{\langle u_0,u_0\rangle_N}
u_0(y)-\frac{\langle h_f^{(0)},\widetilde h_t\rangle_N}{\langle \widetilde h_t,\widetilde h_t\rangle_N}\widetilde h_t(y).
\label{eq:S_toy_defocus_projection}
\end{equation}
The normalized toy-code pair is
\begin{equation}
h_t(y)=\frac{\widetilde h_t(y)}{\|\widetilde h_t\|_N},\quad h_f(y)=\frac{\widetilde h_f(y)}{\|\widetilde h_f\|_N}.
\label{eq:S_normalized_toy_codes}
\end{equation}
This procedure gives the toy codes the same nuisance rejection and normalization convention used for the optimized TRY codes.  Therefore any difference in Fisher information comes from the different code subspaces, not from a different treatment of background or noise.

The toy-code transfer matrix and covariance are computed from the same physical response functions $g_t(y)$ and $g_f(y)$,
\begin{equation}
G^{\rm toy}_{m\mu}=\int dy\,h_m(y)g_\mu(y),\quad
\Sigma^{\rm toy}_{mn}=\int dy\,N(y)h_m(y)h_n(y),
\label{eq:S_toy_transfer_covariance}
\end{equation}
where $m\in\{t,f\}$ and $\mu\in\{t,f\}$. The corresponding coded Fisher matrix is
\begin{equation}
F^{\rm toy}=\left(G^{\rm toy}\right)^{\rm T}
\left(\Sigma^{\rm toy}\right)^{-1}G^{\rm toy}.
\label{eq:S_toy_Fisher}
\end{equation}

For the same physical geometry used in the main text, this gives
\begin{equation}
F^{\rm toy}=\begin{pmatrix}
4.5992\times 10^{-12} & 4.2067\times 10^{-12}\\
4.2067\times 10^{-12} & 4.2061\times 10^{-11}
\end{pmatrix}.\label{eq:S_toy_Fisher_numeric}
\end{equation}
Compared with the full source-resolved Fisher matrix, the retained principal information fractions are
\begin{equation}
\operatorname{eig}\!\left[\left(F^{\rm full}\right)^{-1/2}F^{\rm toy}\left(F^{\rm full}\right)^{-1/2}\right]=\{0.07988,\;0.53729\}.
\label{eq:S_toy_retention_eigenvalues}
\end{equation}
Thus the Gaussian toy-code pair retains only about $8.0\%$ of one principal information direction and $53.7\%$ of the other.

This comparison identifies the main limitation of the toy basis. The true local score functions are not smooth odd/even envelopes. They contain rapid oscillations set by the finite-width Fresnel propagation and by the operating detector position.  The optimized source codes follow those fringe-locked score functions, whereas the toy codes only capture the broad parity structure. As a result, the toy codes project onto the correct qualitative tilt/defocus directions but miss a large fraction of the physically available information.

The conclusion is therefore specific and practical: the Gaussian toy model is useful for intuition, but it is not reliable for designing near-optimal TRY source codes. For the finite-width double-slit system studied here, the codes must be computed from the exact physical Fresnel response.

\section{Finite-slit origin of defocus sensitivity}
This section explains why the even defocus-like perturbation is weak in the narrow-slit limit and becomes visible only when the finite slit width is included.  The effect is not a numerical artifact; it follows directly from the form of the quadratic phase sampled by the two apertures.

The two dimensionless perturbation generators are
\begin{equation}
q_t(x)=\frac{x}{W},\quad q_f(x)=\left(\frac{x}{W}\right)^2,\quad W=\frac{d}{2}.
\label{eq:S_width_generators}
\end{equation}
In the point-slit limit, the two apertures sample only the points $x=\pm W$.  Therefore
\begin{equation}
q_t(+W)=1,\quad q_t(-W)=-1,\label{eq:S_point_tilt}
\end{equation}
whereas
\begin{equation}
q_f(+W)=q_f(-W)=1.\label{eq:S_point_defocus}
\end{equation}
The tilt phase is therefore differential between the two slits, but the defocus phase is common to both slits. In this limit,
\begin{equation}
E(y|\theta_f)\simeq e^{i\theta_f}E_0(y),
\label{eq:S_point_defocus_field}
\end{equation}
so the intensity is unchanged:
\begin{equation}
R(y|\theta_f)=|E(y|\theta_f)|^2\simeq|E_0(y)|^2
=R_0(y).\label{eq:S_point_defocus_intensity}
\end{equation}
Consequently,
\begin{equation}
g_f(y)=\left.\frac{\partial R(y|\theta_f)}{\partial \theta_f}\right|_0\simeq0.
\label{eq:S_point_defocus_score}
\end{equation}
Thus a two-point aperture cannot sense this normalized quadratic phase to first order, because the phase is nearly a global phase.

The same conclusion can be written in a form that is useful for the finite-width calculation. Since
\begin{equation}
-2\,\operatorname{Im}\left[E_0^*(y)E_0(y)\right]=0,
\label{eq:S_common_phase_no_signal}
\end{equation}
the defocus response can be written equivalently as
\begin{equation}
g_f(y)=-2\,\operatorname{Im}\left[E_0^*(y)
\int dx\,K(x,y)\left(q_f(x)-1\right)\right].
\label{eq:S_defocus_centered_score}
\end{equation}
This expression makes clear that the common part of $q_f(x)$ carries no first-order intensity information. Only the variation of $q_f(x)$ across the finite aperture contributes.

To see this variation explicitly, write the coordinate inside each slit as
\begin{equation}
x=sW+u,\quad s=\pm 1,\quad-\frac{a}{2}\le u\le \frac{a}{2}.\label{eq:S_local_slit_coordinate}
\end{equation}
Then
\begin{equation}
q_f(sW+u)=\left(s+\frac{u}{W}\right)^2=1+2s\frac{u}{W}+\left(\frac{u}{W}\right)^2.
\label{eq:S_defocus_expansion}
\end{equation}
The first term is the common phase already removed in Eq.~\eqref{eq:S_defocus_centered_score}. The remaining terms are present only because each slit has finite width. They allow the quadratic phase to be sampled within each opening and convert the defocus-like perturbation into measurable TRY contrast.

We quantified this effect by scanning the slit width while keeping the other parameters fixed:
\begin{equation}
\lambda=633~{\rm nm},\quad L_1=L_2=0.35~{\rm m},\quad
d=500~\mu{\rm m},\quad X_D=-\frac{L_2\lambda}{4d}.
\label{eq:S_width_scan_parameters}
\end{equation}
For each width $a$, we recomputed $R_0(y)$, $g_t(y)$, $g_f(y)$, the noise weight $N(y)=R_0(y)+0.02\max_y R_0(y)$, and the full source-resolved Fisher matrix
\begin{equation}
F^{\rm full}_{\mu\nu}(a)=\int dy\,\frac{g_\mu(y;a)g_\nu(y;a)}{N(y;a)}.
\label{eq:S_width_scan_Fisher}
\end{equation}
As a compact measure of the relative visibility of the two channels, we use
\begin{equation}
\rho(a)=\frac{F^{\rm full}_{ff}(a)}{F^{\rm full}_{tt}(a)}.\label{eq:S_width_ratio}
\end{equation}
Both parameters are dimensionless because of the normalization by $W$, so this ratio compares the local Fisher information for the chosen tilt and defocus parametrization.

\begin{figure}[htbp]
\centering
\includegraphics[width=0.65\linewidth]{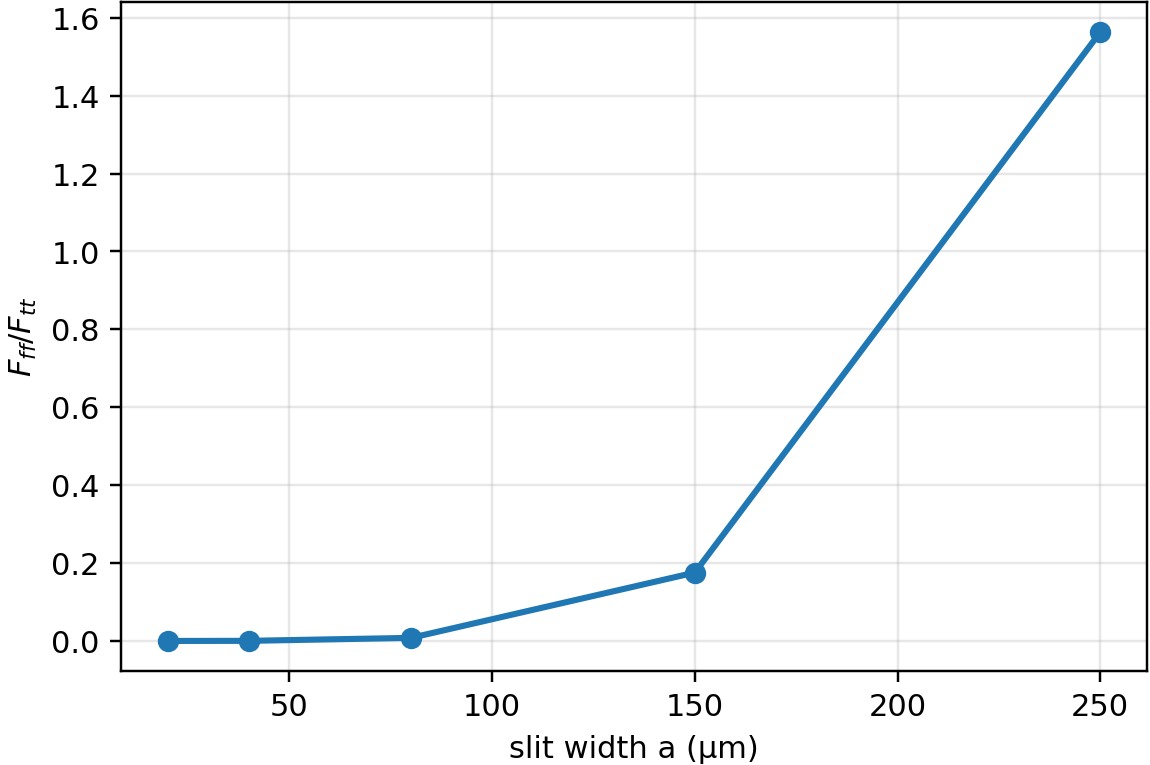}
\caption{Ratio of defocus-to-tilt Fisher information versus slit width $a$. The defocus-like mode is almost invisible in the narrow-slit limit and becomes significant only when the finite slit width is large enough to sample the quadratic phase across each opening.}
\label{fig:widthscan}
\end{figure}

The numerical results are
\begin{table}[h]
\centering
\caption{Relative defocus-to-tilt Fisher information for different slit widths.}
\label{tab:S_width_scan}
\begin{tabular}{c c}
\hline
Slit width $a$ & $\rho(a)=F^{\rm full}_{ff}/F^{\rm full}_{tt}$ \\
\hline
$20~\mu{\rm m}$  & $1.75\times 10^{-5}$ \\
$40~\mu{\rm m}$  & $2.99\times 10^{-4}$ \\
$80~\mu{\rm m}$  & $7.93\times 10^{-3}$ \\
$150~\mu{\rm m}$ & $1.76\times 10^{-1}$ \\
$250~\mu{\rm m}$ & $1.56$ \\
\hline
\end{tabular}
\end{table}
Alternatively, the same trend is shown graphically in Fig.~\ref{fig:widthscan}.
The trend confirms the analytic point above. For very narrow slits, the defocus-like phase is almost indistinguishable from a common phase and its first-order response is strongly suppressed. As the slit width increases, the quadratic phase is sampled across each opening, and the even defocus channel becomes a genuine first-order sensing mode. Therefore, the finite-width Fresnel model is essential not only for quantitative accuracy but also for the existence of appreciable defocus sensitivity in this TRY geometry. In other words, finite slit width is not a cosmetic detail; it is essential for sensing the even defocus-like mode. 

\section{Downstream benchmarks and scope of comparison}
This section clarifies the comparison with downstream strategies. The comparison is local: all Fisher matrices are evaluated for small tilt and defocus perturbations around the same operating point. It is also task-specific: the goal is not full wavefront reconstruction, but estimation of two low-dimensional parameters, $\theta_t$ and $\theta_f$.

For a generic downstream intensity measurement at a transverse coordinate $X$, let the detected field be
\begin{equation}
U(X|\theta_t,\theta_f),\label{eq:S_downstream_field}
\end{equation}
and let
\begin{equation}
I(X|\theta_t,\theta_f)=|U(X|\theta_t,\theta_f)|^2
\label{eq:S_downstream_intensity}
\end{equation}
be the measured intensity distribution in one observation plane. The corresponding local intensity-response functions are
\begin{equation}
h_\mu(X)=\left.\frac{\partial I(X|\theta_t,\theta_f)}
{\partial\theta_\mu}\right|_0,\quad\mu\in\{t,f\}.
\label{eq:S_downstream_scores}
\end{equation}
With a local noise weight $M(X)$, the single-plane direct-intensity Fisher matrix is
\begin{equation}
F_{\mu\nu}^{\rm DI}=\int dX\,\frac{h_\mu(X)h_\nu(X)}{M(X)}.\label{eq:S_DI_Fisher_general}
\end{equation}
This is the downstream analogue of the source-resolved TRY Fisher matrix, except that the measured coordinate is now the downstream image-plane coordinate $X$ rather than the upstream source coordinate $y$.

For a simple near-focus intensity receiver, tilt and defocus do not enter symmetrically.  A small tilt mainly displaces the downstream intensity profile,
\begin{equation}
I(X|\theta_t)\simeq I_0(X-\alpha\theta_t),
\label{eq:S_tilt_shift_model}
\end{equation}
so that
\begin{equation}
h_t(X)\simeq-\alpha\frac{dI_0(X)}{dX}.
\label{eq:S_tilt_shift_score}
\end{equation}
Thus tilt is visible to first order through a centroid or profile shift.

By contrast, a small defocus near the best-focus operating plane is primarily a phase-curvature perturbation. In the idealized local limit where the unperturbed field at the observation plane is chosen real and the defocus perturbation is in optical quadrature, one may write
\begin{equation}
\left.\frac{\partial U}{\partial\theta_f}\right|_0
=iA_f(X)U_0(X),\label{eq:S_defocus_quadrature_model}
\end{equation}
with real $A_f(X)$. The first-order intensity response is then
\begin{equation}
h_f(X)=2\,\operatorname{Re}\left[U_0^*(X)\left.
\frac{\partial U}{\partial\theta_f}\right|_0\right]
=2\,\operatorname{Re}\left[iA_f(X)|U_0(X)|^2\right]
=0.\label{eq:S_defocus_zero_score}
\end{equation}
This is the local reason why a single near-focus intensity plane is a weak receiver for defocus: the defocus perturbation can be predominantly phase-like and therefore invisible to intensity at first order.

In that simplified local setting, the direct-intensity Fisher matrix has the rank-deficient form
\begin{equation}
F^{\rm DI}\simeq
\begin{pmatrix}
F^{\rm DI}_{tt} & 0\\
0 & 0
\end{pmatrix}.\label{eq:S_DI_rank_deficient}
\end{equation}
Equation~\eqref{eq:S_DI_rank_deficient} should not be read as a universal theorem for all downstream measurements. It describes the common single-plane, near-focus, direct-intensity benchmark used for comparison. Defocus can be made first-order observable by adding phase diversity, measuring multiple planes, moving away from the symmetry point, or using a dedicated wavefront or modal analyzer.

An ideal downstream matched-mode sorter provides a stronger benchmark. Suppose a downstream analyzer projects the optical field onto modes that span the same local score subspace associated with $\theta_t$ and $\theta_f$. Then, in the same local Gaussian-noise or shot-noise approximation, its Fisher matrix can attain the full two-parameter Fisher information available in the measured field:
\begin{equation}
F^{\rm sorter}=F^{\rm full}\label{eq:S_sorter_bound}
\end{equation}
when the sorter modes exactly match the relevant local score directions. Therefore, TRY does not claim a stronger information-theoretic bound than an ideal mode sorter. The distinction is architectural. A downstream sorter performs the modal projection after propagation, using a dedicated analyzer. TRY performs the corresponding low-dimensional projection upstream through source coding, while the detector remains fixed.

Full-field downstream wavefront sensing is still more general. Instead of estimating only $\theta_t$ and $\theta_f$, a wavefront sensor usually reconstructs a larger phase function, often written schematically as
\begin{equation}
\phi(x)=\sum_{\ell}a_\ell Z_\ell(x),
\label{eq:S_wavefront_expansion}
\end{equation}
where the $Z_\ell$ may represent Zernike or other modal basis functions. Such a measurement can estimate many aberration coefficients and can therefore solve a broader problem than the one considered here. The price is a higher-dimensional measurement record and, typically, more downstream optical or detector complexity.

The intended comparison can therefore be summarized as follows. Relative to plain single-plane downstream intensity sensing, TRY gives first-order access to both the tilt-like and defocus-like channels in the fixed-detector architecture. Relative to an ideal downstream matched-mode sorter, TRY does not improve the fundamental Fisher bound; instead, it relocates the matched projection from the downstream detection stage to the upstream source basis. Relative to full-field wavefront sensing, TRY is narrower but simpler: it targets a small set of alignment and focus-drift parameters rather than reconstructing an entire wavefront.

\section{Novelty, scope, and limitations of our work}
This section clarifies what is new in the present work and what is not being claimed. The Fisher-information formalism, matched filtering, and nuisance-mode projection used above are standard tools. The new element is their physical realization in a time-reversed Young (TRY) geometry, where the modal projection is implemented by upstream source coding while the detector remains fixed (i.e., through the position-labeled-intensity hybrid correlation channel $R(y)=\avg{\Pi_y I_D}$).

The central physical chain is
\begin{equation}
\text{finite-width TRY Fresnel model}
\;\longrightarrow\;\{g_t(y),g_f(y)\}
\;\longrightarrow\;\{w_t(y),w_f(y)\}
\;\longrightarrow\;F^{\rm TRY}_{2{\rm ch}} .
\label{eq:S_novelty_chain}
\end{equation}
Each step is important. The finite-width Fresnel model determines the exact local response functions $g_t(y)$ and $g_f(y)$. These response functions determine the noise-weighted source codes $w_t(y)$ and $w_f(y)$. The two coded scalar outputs then retain essentially the same local Fisher information as the full source-resolved TRY record for the physical geometry studied here.

The first new point is therefore not the algebraic form of the Fisher matrix, but the physical origin of the score functions. The functions
\begin{equation}
g_t(y)=-2\,\operatorname{Im}\!\left[E_0^*(y)M_t(y)\right],\quad
g_f(y)=-2\,\operatorname{Im}\!\left[E_0^*(y)M_f(y)\right]\label{eq:S_novelty_scores}
\end{equation}
are derived from the exact finite-width double-slit Fresnel integral. They are not assumed to be Gaussian derivatives, Hermite modes, or simple parity functions. Consequently, the optimized source codes are locked to the actual TRY fringe structure.

The second new point is architectural. In a conventional downstream strategy, the field is first propagated to a detection plane and then analyzed by a camera, a wavefront sensor, or a mode sorter. In the present TRY strategy, the low-dimensional projection is imposed upstream in the source basis:
\begin{equation}
S_m=\int dy\,w_m(y)R(y|\theta_t,\theta_f).
\label{eq:S_upstream_projection}
\end{equation}
The detector records only scalar outputs associated with the chosen source codes. Thus, the modal analysis is moved from the downstream detection stage back to the programmable source coordinate.

The third new point is the role of finite slit width. In the point-slit limit, the normalized quadratic phase is common to the two apertures and the defocus-like first-order response nearly vanishes. The finite-width calculation shows that appreciable defocus sensitivity appears only when the quadratic phase is sampled across each slit opening. This is why the physical model gives a result that the Gaussian toy intuition does not capture reliably.

The fourth new point is the quantitative compression result. For the representative geometry studied here,
\begin{equation}
\operatorname{eig}\!\left[\left(F^{\rm full}\right)^{-1/2}F^{\rm TRY}_{2{\rm ch}}
\left(F^{\rm full}\right)^{-1/2}\right]
=\{0.99980958,\;1.00000000\}.
\label{eq:S_novelty_retention}
\end{equation}
Thus, two optimized source-coded channels retain essentially all of the local information contained in the full source-resolved TRY record.  By contrast, the Gaussian odd/even toy-code pair retains only
\begin{equation}
\{0.07988,\;0.53729\}
\label{eq:S_novelty_toy_retention}
\end{equation}
of the two principal information directions for the same physical system. This comparison shows that the exact Fresnel-derived source codes are not a cosmetic refinement; they are necessary for near-lossless compression.

The scope of the result is deliberately limited. We do not claim that TRY surpasses an ideal downstream matched-mode sorter in a fundamental information-theoretic sense. If a downstream analyzer exactly spans the same local score space, it can attain the same Fisher information. The advantage of TRY is instead that the required projection is implemented upstream through source coding while the detector remains fixed.

We also do not claim that TRY replaces full-field wavefront sensing. Full wavefront sensing is designed to reconstruct many aberration modes or a general phase profile. The present scheme targets a narrower task: low-dimensional monitoring of tilt-like and defocus-like drift in a compact fixed-detector architecture.

In this sense, the contribution is specific but practical. The work shows that a finite-width TRY interferometer can be used as an upstream modal sensor: the physical Fresnel response determines the optimal source codes, and those codes convert a distributed source-space response into a small number of fixed-detector channels without appreciable local information loss for the two-parameter task studied here.